# High contrast resonances of the coherent population trapping on sublevels of the ground atomic term

Ersoy Sahin, [1,2,*] Ramiz Hamid,[1] Cengiz Birlikseven,[1] Gönül Özen,[2] and Azad Ch. Izmailov[3]

[1]*National Metrology Institute of Turkey, Gebze, Kocaeli, TURKEY*
[2]*Istanbul Technical University, Faculty of Science and Letters, Engineering Physics Department Maslak, Istanbul, TURKEY*
[3]*Institute of Physics, Azerbaijan National Academy of Sciences, H. Javid av. 33, Baku, Az-1143, AZERBAIJAN*

[*]**Corresponding author**: ersoy.sahin@ume.tubitak.gov.tr

We have detected and analyzed narrow, high contrast coherent population trapping resonances, which appear in transmission of the probe monochromatic light beam under action of the counterpropagating two-frequency laser radiation, on example of the nonclosed three level Λ-system formed by spectral components of the Doppler broadened D2 line of cesium atoms (in the cell with the rarefied Cs vapor). These nontrivial resonances are determined directly by the trapped atomic population on the definite lower level of the Λ-system and may be used in atomic frequency standards, sensitive magnetometers and in ultrahigh resolution laser spectroscopy of atoms and molecules.

At the phenomenon of the coherent population trapping (CPT), a multilevel quantum system subject to decay processes is coherently driven into a superposition state immune from the further excitation, in which the system population is trapped. The CPT is the basis of a number of important applications: ultrahigh resolution spectroscopy, atomic clocks, magnetometry, optical switching, coherent cooling and trapping of atoms and also in some others described, for example in reviews [1-3]. In particular, narrow CPT resonances, detected in absorption of a two-frequency laser radiation (and also in the corresponding induced fluorescence spectrum of a gas medium) on three level atomic Λ-systems, are successfully applied in atomic frequency standards [4] and in high sensitive magnetometers [5]. For these applications, researchers use, mainly, vapors of alkali earth atoms (in particular Cs or Rb), whose ground quantum term consists of two sublevels of the hyperfine structure [6]. Resonance excitation of atoms on the Λ-system scheme is realized by means of the two-frequency radiation from given sublevels (Fig.1). Such Λ-systems are not closed because of presence of channels of the radiative decay of the excited state $|3>$ on some Zeeman sublevels of lower levels which don't interact with the two-frequency radiation [1-3]. Therefore highly narrow CPT resonances, recorded by known methods on the population of the upper level $|3>$ (Fig.1), have a comparatively small contrast on a more wide spectral background of absorption or fluorescence [4].

At the same time, more contrast CPT resonances may appear in dependences of populations of lower long-lived levels $|1>$ and $|2>$ of a nonclosed Λ-system (Fig.1) on the frequency difference $(\omega_2 - \omega_1)$ of the bichromatic laser pumping. Indeed, let us consider interaction of such a system with 2 monochromatic laser fields. Frequencies $\omega_1$ and $\omega_2$ of given fields are close to centers $\Omega_{31}$ and $\Omega_{32}$ of electrodipole transitions $|1> - |3>$ and $|2> - |3>$ respectively (Fig.1). The population of this nonclosed Λ-system will be exhausted at intensification of the two-frequency laser pumping with the exception of a fraction of atoms, which may remain on lower levels $|1>$ and $|2>$ at the following CPT condition [1-3]:

$$|\delta_2 - \delta_1| \leq W \qquad (1)$$

where $\delta_1=(\omega_1 - \Omega_{31})$ and $\delta_2=(\omega_2 - \Omega_{32})$ are detunings of laser frequencies. The width W of the CPT resonance in the expression (1) is determined by intensities of laser fields and by relaxation rates of populations and coherence of quantum states $|1>$ and $|2>$. Under definite conditions, the value W may be much less than homogeneous widths of spectral lines of optical transitions $|1> - |3>$ and $|2> - |3>$ (Fig.1). Given nontrivial CPT resonances may be detected by means of an additional probe radiation resonant to a quantum transition from any lower level $|1>$ or $|2>$. It is important to note that corresponding nonclosed Λ-systems are characteristic not only for many atoms but also for all molecules. In particular, interesting features and possible applications of the CPT phenomenon in a Λ-system of three Zeeman degenerate molecular levels were considered in the theoretical work [7].

In present work we have detected and analyzed nontrivial CPT resonances, which appear in transmission of the probe monochromatic light beam under action of the counterpropagating two-frequency laser radiation on example of the nonclosed Λ-system, formed by spectral components of the $D_2$ line of Cs atoms (Fig.2). Effective natural and Doppler widths of corresponding atomic optical transitions are about 5.3 MHz and 460 MHz (at the room temperature) respectively [6]. The block diagram of our experimental setup is shown in Fig 3. Two independent external cavity diode lasers (ECDL-1 and ECDL-2) were used as monochromatic light sources. The frequency of the ECDL-1 was electronically stabilized on the transmission resonance of the tunable Fabry–Perot interferometer (FPI) using length modulation of FPI and tuned around the transition $6S_{1/2}(F=4) - 6P_{3/2}(F'=3)$ of Cs atoms. The second laser (ECDL-2) frequency was stabilized on the transition $6S_{1/2}(F=3) - 6P_{3/2}(F'=3)$ using the reference Cs atomic cell and the Zeeman modulating technique. ECDL-1 and ECDL-2 laser beams were combined on the beam splitter ($BS_1$) and the resulting bichromatic pumping laser radiation passed through the polarizer ($P_1$) and magnetically shielded cell containing the rarefied Cs vapor at the sufficiently low pressure about 0.1 mPa ($3 \times 10^{10}$ atom/cm$^3$). At the same time, the monochromatic probe laser beam, obtained from the ECDL-2, was sent to this Cs cell in the opposite direction (Fig.3). Given pumping and probe beams were overlapped and had the same linear polarization in the irradiated Cs cell, which was 3 cm long and kept at the temperature of 22 °C. A residual magnetic field inside this cell was less than 10 mGs. Transmission of the monochromatic probe and bichromatic pumping laser beams were detected by photodetectors $PD_2$ and $PD_1$, respectively, and recorded by computers versus the frequency detuning $\delta_1$ of the ECDL-1 at the fixed frequency detuning $\delta_2=0$ of the ECDL-2. The power of the probe beam was kept constant. However we may change the two-frequency pumping beam power and ratio of intensities $I_1$ and $I_2$ of its spectral components, obtained from ECDL-1 and ECDL-2 respectively. The diameter of the bichromatic pumping beam was 5 mm and its total power was about 2.5 mW. In order to avoid visible nonlinear optical effects induced by the probe beam, its power and diameter were essentially smaller: 5 μW and 1 mm respectively.

According to selection rules, at absence of an external magnetic field and orientation of the quantization axis along the same linear polarization of pumping and probe laser beams at our experimental conditions (Fig.3), only optical transitions between Zeeman degenerate Cs levels without change of the magnetic quantum number m are induced [8]. Thus we have 6 nonclosed Λ-systems (corresponding to magnetic numbers m=±1, ±2 and ±3) for two resonant adjacent optical transitions: $6S_{1/2}(F=3) - 6P_{3/2}(F'=3)$ and $6S_{1/2}(F=4) - 6P_{3/2}(F'=3)$ (Fig.2), where CPT resonances are formed. Optical repumping of the population of this resultant Zeeman degenerate Λ-system occurs on three magnetic sublevels of the ground Cs term $6S_{1/2}(F=4, m=\pm4)$ and $6S_{1/2}(F=3, m=0)$, which don't interact with incident pump and probe radiations. Taking into account given features, further we will analyze obtained results on the basis of the simple model of the nonclosed Λ-system (Fig.1), where quantum states $|1>$, $|2>$ and $|3>$ correspond to Cs levels $6S_{1/2}(F=4)$, $6S_{1/2}(F=3)$ and $6P_{3/2}(F'=3)$ (Fig.2). Figs.4a,c present the narrow CPT deep with the center $\delta_1=\delta_2=0$ in transmission of the probe light beam on the resonant transition $|2> - |3>$ (Fig.1). This resonance is caused directly by the trapping of an atomic population fraction of the lower level $|2>$ in the nonclosed Λ-system (Fig.1) because of a negligible population of its excited state $|3>$ at the CPT condition (1). Given deep appears on the background of the essentially more wide and less contrast sub-Doppler (lamb) deep [8], which is caused by the velocity selected optical repumping of a population fraction from the level $|1>$ to $|2>$ (Fig.1) because of the light induced transition $|1> \leftrightarrow |3>$ and the following radiative decay on the channel $|3> \rightarrow |2>$. Unlike Figs.4a,c, the well-known narrow CPT peak in the transmission of the corresponding two-frequency laser pumping has the less amplitude in comparison with its sub-Doppler background (Fig.4b,d). Detected CPT resonances have following approximate contrasts C (in % with respect to the total recorded background) and widths W (on their half-heights), calculated according to definitions of the paper [9]: 8% and 4.166 MHz (Fig.4a), 2.3 % and 3.968 MHz (Fig.4b), 3.5% and 4.028 MHz (Fig.4c), 1 % and 3.043 MHz (Fig.4d). Given values C and W were less for CPT resonances in Figs.4c,d than in corresponding Figs.4a,b, which were obtained at the same pumping intensity component $I_2$ but for the essentially different intensity of another component $I_1=0.1I_2$ (Figs.4c,d) and $I_1=I_2$ (Figs.4a,b). We note that the CPT resonance in the transmission of the probe beam (Figs.4a,c) has the close width but essentially more contrast in comparison with the corresponding well-known CPT resonance in transmission of the two-frequency pumping radiation (Figs.4b,d). Results of our calculations, carried out on the basis of density matrix equations in the model of nondegenerate quantum levels of the nonclosed Λ-system (Fig.1), are in good qualitative agreement with corresponding experimental data presented in Fig.4. More detailed experimental and theoretical research of given CPT resonances versus parameters of the two-frequency pumping beam (in particular on its intensities $I_1$, $I_2$) will be carried out in our following works. It is necessary to note that, even a small angle (~ a few degrees) between linear polarizations of pumping and probe laser beams or a weak external magnetic field (~1 Gs) in the Cs cell (Fig.3) lead to essential decrease of contrasts of given recorded CPT resonances. This is caused by difference of the real system of Zeeman degenerate levels (Fig.2) from the considered theoretical model of the Λ-system (Fig.1).

Thus, on the basis of the elaborated method, we have detected and analyzed new narrow, high contrast CPT resonances characteristic for a nonclosed atomic (or molecular) three level Λ-system (Fig.1). These CPT resonances are determined, mainly, by the trapped population on the definite lower level of the Λ-system, from which the resonant optical transition is induced by the weak probe radiation. Given resonances may be used in atomic frequency standards and sensitive magnetometers based on the CPT phenomenon and also in ultrahigh resolution laser spectroscopy of atoms and molecules.

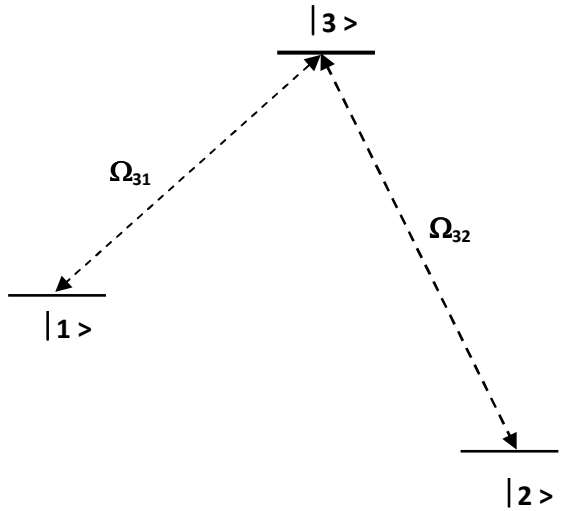

Fig.1. The $\Lambda$-system of optical transitions $|1> - |3>$ and $|2> - |3>$ (with central frequencies $\Omega_{31}$ and $\Omega_{32}$) between the excited level $|3>$ and long lived states $|1>$ and $|2>$.

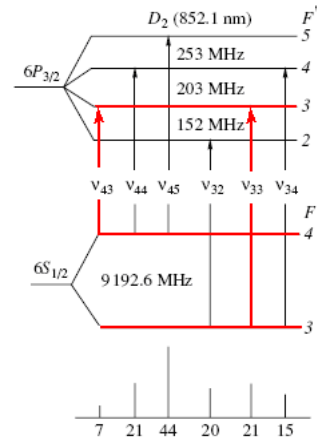

Fig.2. Energy level scheme of the Cs $D_2$ line. The relative oscillator strengths of lines, representing hyperfine transitions, are given on the bottom of this figure.

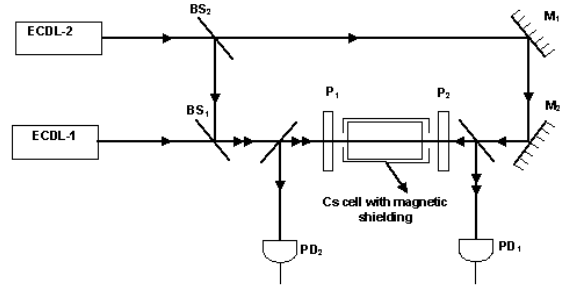

Fig.3. Scheme of our experimental setup, which includes two external cavity diode lasers (ECDL-1, ECDL-2), beam splitters ($BS_1$, $BS_2$), mirrors ($M_1$, $M_2$), polarizers ($P_1$, $P_2$), photodiodes ($PD_1$, $PD_2$) and the Cs cell with the magnetic shielding.

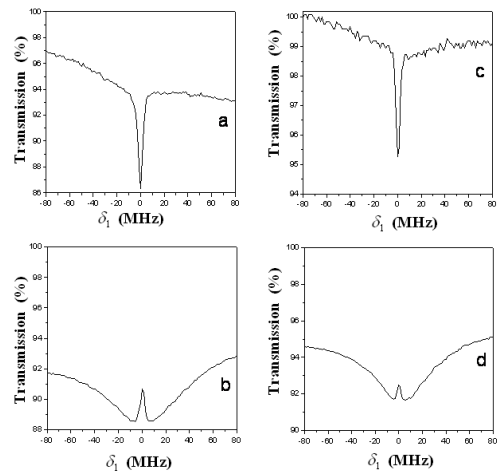

Fig.4. Detected CPT resonances in transmission of the probe beam (a,c) and in transmission of the corresponding two-frequency pumping radiation (b,d) versus the frequency detuning $\delta_1$ at the fixed detuning $\delta_2=0$, when $I_1=I_2$ (a,b) and $I_1=0.1 I_2$ (c,d), $I_2=$ 6.4 mW/cm$^2$ and $I_{probe}=$ 0.64 mW/cm$^2$.